\documentclass[twocolumn,showpacs,preprintnumbers,amsmath,amssymb]{revtex4}

\usepackage{color,epsfig,graphicx,enumerate,bm}
\usepackage{bbold}
\usepackage{bbm}
\usepackage{latexsym}
\usepackage{amsmath}
\usepackage{amssymb}
\usepackage{amsfonts}
\usepackage{graphicx}
\usepackage{epstopdf}
\newcommand{\mi}{\mathrm{i}}
\newcommand{\rmd}{\mathrm{d}}

\newcommand{\ag}{ {\alpha\gamma} }

\newcommand{\dubav}[1]{\langle \! \langle #1 \rangle \! \rangle}

\makeatletter
\def\overbracket{\@ifnextchar [ {\@overbracket} {\@overbracket
[\@bracketheight]}}
\def\@overbracket[#1]{\@ifnextchar [ {\@over@bracket[#1]}
{\@over@bracket[#1][0.3em]}}
\def\@over@bracket[#1][#2]#3{%\message {Overbracket: #1,#2,#3}
\mathop {\vbox {\m@th \ialign {##\crcr \noalign {\kern 3\p@
\nointerlineskip }\downbracketfill {#1}{#2}
                              \crcr \noalign {\kern 3\p@ }
                              \crcr  $\hfil \displaystyle {#3}\hfil $%
                              \crcr} }}\limits}
\def\downbracketfill#1#2{$\m@th \setbox \z@ \hbox {$\braceld$}
                  \edef\@bracketheight{\the\ht\z@}\downbracketend{#1}{#2}
                  \leaders \vrule \@height #1 \@depth \z@ \hfill
                  \leaders \vrule \@height #1 \@depth \z@ \hfill
\downbracketend{#1}{#2}$}
\def\downbracketend#1#2{\vrule depth #2 width #1\relax}
\makeatletter
\def\underbracket{%
  \@ifnextchar [ %
    {\@underbracket}%
    {\@underbracket [\@bracketheight]}}
\def\@underbracket[#1]{%
  \@ifnextchar [ %
    {\@under@bracket[#1]}%
    {\@under@bracket[#1][0.4em]}}
\def\@under@bracket[#1][#2]#3{%\message {Underbracket: #1,#2,#3}
  \mathop {%
    \vtop {%
      \m@th \ialign {%
	##\crcr $\hfil \displaystyle {#3}\hfil $%
       \crcr \noalign %
       {\kern 3\p@ \nointerlineskip }%
	\upbracketfill {#1}{#2}
       \crcr \noalign %
       {\kern 3\p@ }%
     }%
   }%
  }%
  \limits%
}
\def\upbracketfill#1#2{%
  $\m@th \setbox \z@ \hbox {$\braceld$}
  \edef\@bracketheight{\the\ht\z@}\bracketend{#1}{#2}
  \leaders \vrule \@height #1 \@depth \z@ \hfill 
  \leaders \vrule \@height #1 \@depth \z@ \hfill%
  \bracketend{#1}{#2}$%
}
\def\bracketend#1#2{\vrule height #2 width #1\relax}
\begin{document}

%\preprint{APS/123-QED}

\title{Optimal experiment design revisited: fair, precise and minimal tomography.}% Force line breaks with \\

\author{J.~Nunn}
\email{j.nunn1@physics.ox.ac.uk} 
\affiliation{Clarendon Laboratory, University of Oxford, Parks Road,
Oxford OX1 3PU, United Kingdom}
\author{B.~J.~Smith, G.~Puentes}
\affiliation{Clarendon Laboratory, University of Oxford, Parks Road,
Oxford OX1 3PU, United Kingdom}
\author{J.~S.~Lundeen}
\affiliation{National Research Council of Canada, Ottawa, Ontario K1A 0R6, Canada}
\author{I.~A.~Walmsley}
\affiliation{Clarendon Laboratory, University of Oxford, Parks Road,
Oxford OX1 3PU, United Kingdom}

\date{\today}% It is always \today, today,
             %  but any date may be explicitly specified

\begin{abstract}
Given an experimental set-up and a fixed number of measurements, how should one take data in order to optimally reconstruct the state of a quantum system? The problem of optimal experiment design (OED) for quantum state tomography was first broached by Kosut \emph{et al}. \cite{Kosut:2004pi}. Here we provide efficient numerical algorithms for finding the optimal design, and analytic results for the case of `minimal tomography'. We also introduce the \emph{average OED}, which is independent of the state to be reconstructed, and the \emph{optimal design for tomography} (ODT), which minimizes tomographic bias. We find that these two designs are generally similar. Monte-Carlo simulations confirm the utility of our results for qubits. Finally, we adapt our approach to deal with constrained techniques such as maximum likelihood estimation. We find that these are less amenable to optimization than cruder reconstruction methods, such as linear inversion.
\end{abstract}


%\keywords{Suggested keywords}%Use showkeys class option if keyword
%display desired

\maketitle
\section{Introduction}
Quantum state tomography \cite{Nielsen:2004kl,Kosut:2004pi,Hradil:2006ud,Paris:2004kx,ifmmode-Relse-Rfiehaifmmode-celse-cfiek:2004rt} is an indispensable tool in quantum information processing, being essential for the characterization of quantum sources \cite{Puentes:2009rw,Lvovsky:2009ys,Huisman:rc}, gates \cite{OBrien:2004kx,Riebe:2006jt}, processes \cite{Branderhorst:2007fr,Modi:2009qd,Yu:2009hl,Mohseni:2008qr} and measurements \cite{Lundeen:2008hc,Audenaert:2009sp}. The reconstruction of quantum states is, however, essentially a classical problem --- that of estimating the parameters of a density matrix from a data set. The classical theory of multiple parameter estimation is well developed \cite{Cramer:1946fq,Radhakrishna-Rao:1945by,Bard:1974rp,Ollila:2008if,Van-den-Bos:1994ef,Jagannatham:2004fe,Fessler:1996cl,Marzetta:1993hb,Matson:2006vf,Stoica:1998uk,Gorman:1990eq}, and so a considerable amount is known about the precision that can be achieved with quantum tomography \cite{Braunstein:1992gb,Kosut:2004pi,Young:2009ca,Demkowicz-Dobrzanski:2009ys,Dorner:2006sy}.

In this paper we are concerned with designing an experiment so as to optimize this precision. Choosing the right set of measurements is of paramount importance, and the optimal measurements for tomography are now known \cite{Kurzynski:2009oq,Wootters:1989cj,Paterek:2002om,Brierley:2009wu,Weigert:2008fb,Adamson:2008df,Vianna:2008dz}. But it is not always possible to implement them: more often than not, technical constraints permit only a non-ideal set of measurements \cite{Burgh:2008xu}. Given such a set, and a finite time in which to acquire data, one encounters the question ``How much time should be spent on each measurement, so as to perform the best tomographic inversion?''. This is the problem of optimal experiment design --- OED for short. We will see that a judicious design can significantly improve the performance of tomographic reconstruction. Rather paradoxically, however, the OED generally \emph{depends} on the state we wish to reconstruct, so that one cannot find the OED if one is completely ignorant of the quantum state. In this paper we introduce two alternative approaches to experiment design that do not suffer from this state dependence.

The paper is structured as follows. In Section \ref{section:background} we review the theory of multi-parameter estimation, and we introduce the notation to be used subsequently. In Section \ref{section:algorithm} we will show how to find the OED quickly using standard numerical techniques. We then move on in Section \ref{section:analytic} to derive an analytic formula for the OED, which holds in the case that the quantum state is not overdetermined by the available data. Sections \ref{section:average} and \ref{section:odt} address the problem of state tomography when no prior knowledge of the true quantum state exists. That is, when one really has no idea at all what quantum state we expect to find. On the one hand, we may decide that we should design our experiment so that, on average, whatever the true state, the tomography is as precise as possible. This we call the average OED. On the other hand, we might want our tomography to be as `fair' as possible, so that the precision is as close as possible to being independent of the true state. We call this the optimal design for tomography --- ODT. We show that --- fortunately --- the tradeoff between precision and fairness is rather small. In Section \ref{section:monte_carlo} we present the results of Monte-Carlo simulations to corroborate our predictions. Finally in Section \ref{section:cholesky} we consider adapting our results to the case of constrained estimators, such as maximum likelihood estimation: we conclude that our optimizations still apply for such estimators, even though the improvement is less marked than for unconstrained tomography. Section \ref{section:conclusion} wraps up the paper with some concluding remarks.

\section{Background}
\label{section:background}
\subsection{Bloch representation}
We suppose that we are given an ensemble of identical $N$-dimensional quantum systems, all of which are prepared in the same way. Quantum state tomography amounts to estimating $2N^2$ real numbers comprising the elements of $\rho$, the complex-valued $N\times N$ density matrix describing the ensemble. But not all of these numbers are independent of one another, because $\rho$ must always be Hermitian, with unit trace and positive eigenvalues \cite{Nielsen:2004kl}. The first two conditions reduce the number of independent real parameters in the density matrix down to $N^2-1$. An equivalent description of the quantum system is then given by the \emph{Bloch vector} $\bm{r}$, which is a real column vector whose $N^2-1$ elements uniquely determine $\rho$, according to the relation
\begin{equation}
\label{rho1}
\rho = \frac{1}{N}I + \bm{r}.\bm{\sigma},
\end{equation}
where $I$ is the $N\times N$ identity matrix. Here $\bm{\sigma}$ is an $N^2-1$ dimensional vector whose elements are $N\times N$ matrices that form an orthonormal basis for the space of traceless Hermitian operators \cite{Harriman:1967kx,Paris:2004kx}. That is to say,
\begin{equation}
\label{sigmas}
\textrm{tr}\left\{\sigma_j\right\}=0,\qquad \textrm{and}\qquad \textrm{tr}\left\{\sigma_j\sigma_k\right\}=\delta_{jk}.
\end{equation}
Any set of matrices satisfying these conditions will suffice for constructing $\bm{\sigma}$. For $N=2$, the three elements of $\bm{\sigma}$ are conventionally taken to be the familiar Pauli matrices, and then $\bm{r}$ is the standard three dimensional vector that describes the state of a qubit in the Bloch sphere \cite{Nielsen:2004kl}. The vector $\bm{r}$ is known as the \emph{Bloch} or \emph{Fano} representation of $\rho$; its usefulness becomes clear when we consider the way it enters into the calculation of measurement statistics.

We will perform our tomographic experiment by sending multiple copies of the state $\rho$ into our apparatus. The apparatus can be configured in one of $M$ different ways, so that there are $M$ different possible measurements we can make (for instance, we could measure the three Cartesian components of a spin, and $M=3$ in this case). Let $\gamma$ denote the particular measurement we are making. For each $\gamma$, there are $n_\gamma$ different possible measurement outcomes (in our spin example, $n_\gamma = 2S+1$ for all the measurements, where $S$ is the total spin quantum number). Associated to each measurement outcome $\alpha$ is a so-called \emph{positive operator-valued measure element} (POVM element) \cite{Nielsen:2004kl}, which is a Hermitian, positive $N\times N$ matrix $\Pi_\ag$, such that the probability $p_\ag$ of obtaining the outcome $\alpha$ in configuration $\gamma$ is given by\footnote{Much of our notation is plagiarized from the paper by Kosut \emph{et al}. \cite{Kosut:2004pi}, which first introduced the idea of OED for quantum tomography.}
\begin{equation}
\label{prob1}
p_\ag = \textrm{tr}\left\{\Pi_\ag\rho\right\}.
\end{equation}
The POVM elements contain the physics of our tomography set-up. All noise and detector inefficiency can be incorporated into them, so that once they are fixed, the only issues we must deal with are statistical. Like the density matrix, they are Hermitian, and can also be written in the Bloch representation:
\begin{equation}
\label{POVM1}
\Pi_\ag = c_\ag I + \bm{a}_\ag.\bm{\sigma},
\end{equation}
where $c_\ag$ is a real number, and $\bm{a}_\ag$ is a real ($N^2-1$)-dimensional vector. For each measurement configuration $\gamma$, the probabilities $p_\ag$ must sum to unity, and this imposes the sum constraint $\sum_{\alpha=1}^{n_\gamma} \Pi_\ag = I$. In the Bloch representation, we must have, accordingly,
\begin{equation}
\label{sum_constraints}
\sum_{\alpha=1}^{n_\gamma} c_\ag=1, \qquad \textrm{and}\qquad \sum_{\alpha=1}^{n_\gamma} \bm{a}_\ag = 0,
\end{equation}
 for all values of $\gamma$. Now, substituting Eq.~(\ref{POVM1}) into Eq.~(\ref{prob1}) we obtain the simple affine relation
\begin{equation}
\label{prob2}
p_\ag = c_\ag + \bm{a}_\ag.\bm{r},
\end{equation}
which connects the measurement statistics with the Bloch vector. The probabilities can be collected into an $n_\mathrm{tot}$-dimensional vector $\bm{p}$, where
\begin{equation}
\label{n_tot_def}
n_\mathrm{tot} = \sum_{\gamma=1}^M n_\gamma
\end{equation}
is the total number of outcomes our experiment can produce, across all measurement configurations. Then the statistics are summarized by the matrix equation
\begin{equation}
\label{prob3}
\bm{p} = \bm{c} + A\bm{r},
\end{equation}
where $\bm{c}$ is a real vector whose elements are the $c_\ag$, and where $A$ is the $n_\mathrm{tot}\times (N^2-1)$ matrix whose rows are given by the vectorized POVM elements,
\begin{equation}
\label{Adef}
A = \left(\begin{array}{c} \overbracket[.8 pt]{\underbracket[.8 pt]{\;\phantom{\cdots\cdots.}\;\; \bm{a}_{11}^\mathsf{T}\;\; \phantom{\cdots\cdots.}\;\;}} \\
\overbracket[.8 pt]{\underbracket[.8 pt]{\;\phantom{\cdots\cdots.}\;\; \bm{a}_{21}^\mathsf{T}\;\; \phantom{\cdots\cdots.}\;\;}} \\
\vdots  \\
\overbracket[.8 pt]{\underbracket[.8 pt]{\;\phantom{.\cdots\cdots}\;\; \bm{a}_\ag^\mathsf{T}\;\; \phantom{\cdots\cdots.}\;\;}} \\
 \vdots  \\
\overbracket[.8 pt]{\underbracket[.8 pt]{\;\phantom{\cdots\cdots}\;\; \bm{a}_{n_M M}^\mathsf{T}\;\; \phantom{\cdots\cdots}\;\;}} \end{array}
\right).
\end{equation}
This notation uses no more matrix elements than necessary, unlike using $\rho$ and $\Pi_{\alpha\gamma}$ in (\ref{prob1}). As is often the case, the notational compactness of matrices and vectors will prove invaluable in teasing out the structure of the optimizations that follow.

\subsection{Cramer-Rao bound}
Suppose that we use a total of $N_\mathrm{tot}$ copies of our quantum system (note that $N_\mathrm{tot}$ is different to $n_\mathrm{tot}$, the latter being the total number of possible outcomes). We measure $N_\gamma$ times in each of the $M$ configurations, so that we have
\begin{equation}
\label{Ntot}
N_\mathrm{tot}=\sum_{\gamma=1}^M N_\gamma.
\end{equation}
At the end of the experiment, we are left with an $N_\mathrm{tot}$-dimensional data vector $\bm{n}$ whose elements $n_\ag$ are the number of times the outcome $\alpha$ was observed in configuration $\gamma$, so that $\sum_{\alpha=1}^{n_\gamma} n_\ag = N_\gamma$. We must process $\bm{n}$ somehow to produce an estimate of the true state $\bm{r}$. The probability of obtaining the data vector $\bm{n}$ is given (up to an unimportant combinatorial factor) by the likelihood function $p$,
\begin{equation}
\label{prob_data}
p\left(\bm{n}|\bm{r}\right) = \prod_\ag p_\ag^{n_\ag}. 
\end{equation}
We will see below that the sensitivity of this likelihood function to $\bm{r}$, through Eq.~(\ref{prob2}), determines the precision of our tomographic inversion. The OED is the set of numbers $\left\{N_\gamma\right\}$ which makes our estimate of $\bm{r}$ as precise as possible.

Our approach to finding the OED follows that presented in \cite{Kosut:2004pi}, in which we seek to minimize the \emph{Cramer-Rao bound} \cite{Cramer:1946fq,Radhakrishna-Rao:1945by,Bard:1974rp}. This is a classical information-theoretic lower bound on the mean squared error in the estimate of $\bm{r}$, which does not depend on the algorithm used to extract that estimate, as long as the estimate is \emph{unbiased} (We will discuss certain biased estimators in Section \ref{section:cholesky}). The Cramer-Rao bound is defined in terms of the \emph{Fisher information matrix}\footnote{We have scaled the matrix by $N_\mathrm{tot}$, so that strictly speaking we are dealing with the Fisher information \emph{per datum}. The use of this `normalized' matrix will avoid notational clutter elsewhere.}
\begin{equation}
\label{fisher}
F = \frac{1}{N_\mathrm{tot}}\left\langle \bm{\Delta} \bm{\Delta}^\mathsf{T}\right\rangle,
\end{equation}
where $\langle . \rangle$ indicates an ensemble average and the symbol $\mathsf{T}$ denotes the transpose. The vector $\bm{\Delta}= \bm{\nabla}_{\bm{r}}\mathcal{L}$ is the gradient of the \emph{log-likelihood function}
\begin{equation}
\label{loglik}
\mathcal{L} = \ln p\left(\bm{n}|\bm{r}\right) = \sum_\ag n_\ag \ln p_\ag.
\end{equation}
One might expect that `informative' measurements are those associated with a greater sensitivity of $p$ to $\bm{r}$, and indeed with these definitions, a strong $\bm{r}$-dependence of $p$ contributes to the magnitude of the Fisher information $F$.
Substituting Eq.~(\ref{prob2}) into Eq.~(\ref{loglik}), we obtain $\bm{\Delta} = \sum_\ag n_\ag \bm{a}_\ag/p_\ag$, so that
\begin{equation}
\label{Fish_become}
F = \frac{1}{N_\mathrm{tot}}\sum_\ag \sum_{\beta \delta} \frac{\left\langle n_\ag n_{\beta \delta}\right\rangle}{p_\ag p_{\beta \delta}}\bm{a}_\ag \bm{a}_{\beta \delta}^\mathsf{T}.
\end{equation}
There are no correlations between different measurement configurations; within the same configuration, we can use the standard result for multinomial distributions, so we have
\begin{equation}
\label{corr}
\langle n_\ag n_{\beta \delta}\rangle = \delta_{\gamma \delta} \left[ N_\gamma\left(N_\gamma-1\right)p_\ag p_{\beta \delta} + N_\gamma \delta_{\alpha \beta} p_\ag \right].
\end{equation}
Only the last term contributes (the other terms vanish due to the second condition in Eq.~(\ref{sum_constraints}) \cite{Paris:2004kx}). Finally, we are left with the expression
\begin{equation}
\label{fisher2}
F = \frac{1}{N_\mathrm{tot}}\sum_\ag \frac{N_\gamma}{p_\ag}\bm{a}_\ag\bm{a}_\ag^\mathsf{T}.
\end{equation}
The inverse of this matrix is a lower bound on the covariance matrix of our reconstructed state, known as the Cramer-Rao bound. That is, if we estimate the state to be $\hat{\bm{r}}$, then the matrix $N_\mathrm{tot}\textrm{Cov}\left(\bm{r},\hat{\bm{r}}\right) - F^{-1}$ has positive eigenvalues. In particular, taking the trace yields the condition
\begin{equation}
\label{norm_cond}
\left \langle|\bm{r}-\hat{\bm{r}}|^2 \right \rangle \geq B/N_\mathrm{tot},
\end{equation}
where
\begin{equation}
\label{B_def}
B=\textrm{tr}\left\{F^{-1}\right\}.
\end{equation}
The mean squared error in our reconstruction is therefore bounded by the quantity $B$, which we will henceforth refer to
as `the Cramer-Rao bound', or the `CRB' for short. It has a clear operational meaning as the best mean squared error achievable by our tomography experiment (scaled by $N_\mathrm{tot}$). Furthermore, as the simulations in Section \ref{section:monte_carlo} demonstrate, the CRB represents a tight bound  --- in the limit of a large number of measurements, the achieved precision (\emph{i.e.} mean-squared error) for unbiased tomography is well-described by $B/N_\mathrm{tot}$.
 
\section{Numerical optimization}
\label{section:algorithm}
We are now ready to give a more formal rendering of the problem at hand. Our aim in finding the OED is to discover the numbers $\left\{N_\gamma \right\}$ which minimize $B$, subject to the constraint in Eq.~(\ref{Ntot}). Of course the $N_\gamma$ must be integers, since one cannot perform an experiment a fractional number of times! But the solution of this problem is intractable, being combinatorial in nature. We will instead consider what the authors of \cite{Kosut:2004pi} called the \emph{relaxed} problem, where we allow `fractional experiments'. We define the real, positive $M$-dimensional vector $\bm{\lambda}$ with elements $\lambda_\gamma$ such that
\begin{equation}
\label{sum_lam}
\sum_\gamma \lambda_\gamma = 1.
\end{equation}
The $\lambda$'s are the `weights' representing the experiment design, so that in the limit $N_\mathrm{tot}\longrightarrow \infty$ of a large number of measurements, we have
\begin{equation}
\label{Ngamma}
N_\gamma \longrightarrow N_\mathrm{tot}\lambda_\gamma.
\end{equation}
The experiment design $\bm{\lambda}$ is only asymptotically correct, but as our simulations in Section \ref{section:monte_carlo} show, the optimization of $\bm{\lambda}$ produces results which are beneficial for finite data sets with a wide range of sizes: one simply rounds the right hand side of (\ref{Ngamma}) to the nearest integer to obtain $N_\gamma$. The Fisher information becomes
\begin{equation}
\label{fisher3}
F = \sum_\ag \frac{\lambda_\gamma}{p_\ag}\bm{a}_\ag \bm{a}_\ag^\mathsf{T}.\end{equation}
In \cite{Kosut:2004pi}, similar expressions were derived, and a numerical convex optimization routine was invoked to find the vector $\bm{\lambda}$ which minimized $B=\textrm{tr}\left(F^{-1}\right)$, subject to the normalization constraint in Eq.~(\ref{sum_lam}). But this problem does not require specialized convex optimization software. The optimization can be performed quickly using standard gradient-ascent algorithms, since the constraint in Eq.~(\ref{sum_lam}) can be incorporated using a Lagrange multiplier. We define the \emph{cost function}
\begin{equation}
\label{cost1}
J = B + \eta\left(\sum_\gamma \lambda_\gamma -1\right),
\end{equation}
where $\eta$ is a real Lagrange multiplier that imposes the normalization constraint on $\bm{\lambda}$.
The OED is the vector $\bm{\lambda}^\mathrm{OED}$ such that the cost function is rendered stationary,
\begin{equation}
\label{stationary}
\left. \bm{\nabla}_{\bm{\lambda}} J \right|_{\bm{\lambda}=\bm{\lambda}^\mathrm{OED}}=0.
\end{equation}
To proceed further, we need to differentiate $B$. This is most easily done by re-writing the Fisher information as a product of matrices, as follows.
\begin{equation}
\label{fisher_prod}
F = A^\mathsf{T}\Lambda P^{-1}A,
\end{equation}
where $P=\textrm{diag}\left(\bm{p}\right)$ and
\begin{equation}
\label{Lam_def}
\Lambda = \textrm{diag}\left(\left[\lambda_1 \mathbb{1}_{n_1}^\mathsf{T},\ldots,\lambda_\gamma \mathbb{1}_{n_\gamma}^\mathsf{T},\ldots,\lambda_M \mathbb{1}_{n_M}^\mathsf{T}\right]^\mathsf{T}\right)
\end{equation}
are both $n_\mathrm{tot}\times n_\mathrm{tot}$ diagonal matrices, and where $A$ was defined in Eq.~(\ref{Adef}). Here we used $\mathbb{1}_m$ to denote the $m$-dimensional column vector whose elements are all equal to $1$. With the expression in Eq.~(\ref{fisher_prod}) in hand, it is easy to differentiate the cost function \cite{Petersen:2006gf}. We obtain
\begin{equation}
\label{DiffJ}
\bm{\nabla}_{\bm{\lambda}} J = \eta \mathbb{1}_M - \textrm{diag}\left\{\textrm{tr}_\alpha \left[P^{-1}AF^{-2}A^\mathsf{T}\right]\right\},
\end{equation}
where $\left(\textrm{tr}_\alpha \left[X\right]\right)_{\gamma,\delta} = \sum_\alpha X_{\alpha \gamma,\alpha \delta}$ indicates a partial trace over the measurement outcomes. The OED is found by minimizing the norm of $\bm{\nabla}_{\bm{\lambda}}J$, which can be done efficiently using, for example, Matlab's \texttt{lsqnonlin} routine. Convergence is accelerated by providing the algorithm with an expression for the Hessian $H$ of $J$ --- the matrix containing the gradients of $\bm{\nabla}_{\bm{\lambda}}J$:
\begin{eqnarray}
\nonumber
H_{\delta\gamma} &=& \partial_{\lambda_\delta} \partial_{\lambda_\gamma} J\\
\label{hessian} &=& \mathrm{tr}\left[ \left(\partial_{\lambda_\gamma}\Lambda\right) P^{-1}AK_\delta A^\mathsf{T}\right],
\end{eqnarray}
where
\begin{equation}
\label{Kdef}
K_\delta=\left\{F,F^{-1}A^\mathsf{T}P^{-1}\left(\partial_{\lambda_\delta}\Lambda\right)AF^{-1}\right\},
\end{equation}
with $\left\{. , . \right\}$ denoting the anti-commutator. The matrix $\partial_{\lambda_\gamma}\Lambda$ contains the vector $\mathbb{1}_{n_\gamma}$ along the diagonal of its $\gamma^\mathrm{th}$ sub-block, and zeros everywhere else. To proceed with the optimization, we first pick an initial `guess' for $\bm{\lambda}$. We then perform a quick line-search optimization to find the value of $\eta$ that minimizes $J$, given our guess. We then feed $\bm{\lambda}$ and $\eta$ into \texttt{lsqnonlin}.

In Figure \ref{figure:compare_Kosut} below, we plot the OED predicted using the above procedure, for a model polarimetric experiment introduced by Kosut \emph{et al}. \cite{Kosut:2004pi} (see Section 2.4, and in particular the top panel of Figure 3 therein). We will not discuss the various parameters involved; we simply comment that the agreement between our results and those in \cite{Kosut:2004pi} is excellent. The advantage of the method presented here is that no dedicated convex optimization packages are required. The optimization is therefore easier to implement; only standard numerical tools (any `conjugate gradients' algorithm will perform well) are required. Furthermore, the provision of the Hessian makes for very rapid convergence. In Section \ref{section:average}, we will see that this numerical method can also be employed to find the \emph{average} OED.
\begin{figure}[h]
\begin{center}
\includegraphics[width=\columnwidth]{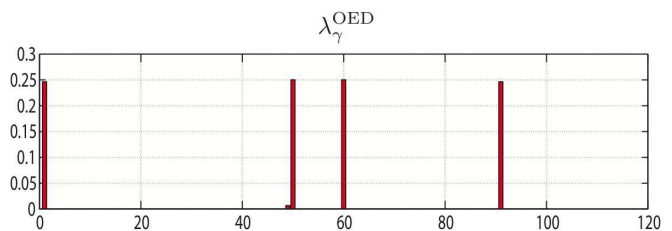}
\caption{The OED for a simple polarization tomography experiment (described in \cite{Kosut:2004pi}). This result is to be compared with the top-panel of Figure 3 in \cite{Kosut:2004pi}. The optimization we use is simple and fast, requiring no specialist software.}
\label{figure:compare_Kosut}
\end{center}
\end{figure}

\section{Minimal OED}
\label{section:analytic}
In certain circumstances, we can do better than the above numerical optimization: we can write down an analytic expression for the OED. This is possible if we are able to invert the Fisher information matrix $F$ `by hand'. The form of Eq.~(\ref{fisher_prod}) is suggestive. If we can take the inverse of the matrices $A$, $P^{-1}$ and $\Lambda$, we can multiply them to form $F^{-1}$. But in general, $A$ is not invertible. First, because only square matrices have inverses, and it may be that $n_\mathrm{tot}\neq N^2-1$, and second, because even if $A$ is square, it must be of full rank, in order to be invertible. That is, none of its rows can be linearly dependent upon any of the others. But the rows of $A$ are given by the $\bm{a}_\ag^\mathsf{T}$, which are \emph{always} linearly dependent upon each other because of the second condition in Eq.~(\ref{sum_constraints}) --- ultimately because of the conservation of probability. This latter issue is potentially fatal, but fortunately it is possible to re-write the Fisher information in terms of full-rank matrices. To see how, note that from Eq.~(\ref{sum_constraints}), we can re-express the last POVM element in each configuration in terms of the other elements,
\begin{equation}
\label{last_a}
\bm{a}_{n_\gamma \gamma} = -\sum_{\alpha'=1}^{\widetilde{n}_\gamma}\bm{a}_{\alpha'\gamma},
\end{equation}
where $\widetilde{n}_\gamma = n_\gamma-1$ is the number of \emph{independent} measurement outcomes associated with configuration $\gamma$. In subsequent calculations we will always use primed indices, such as $\alpha'$, to enumerate these independent outcomes (\emph{i.e.} omitting the last outcome with $\alpha = n_\gamma$).
Substituting Eq.~(\ref{last_a}) into Eq.~(\ref{fisher3}), we obtain
\begin{equation}
\label{}
F = \sum_\gamma \lambda_\gamma \left(\sum_{\alpha'}\frac{\bm{a}_{\alpha'\gamma}\bm{a}_{\alpha'\gamma}^\mathsf{T}}{p_{\alpha'\gamma}} + \frac{1}{p_{n_\gamma \gamma}}\sum_{\alpha'\beta'}\bm{a}_{\alpha'\gamma}\bm{a}_{\beta'\gamma}^\mathsf{T}\right).
\end{equation}
Some re-definitions then afford a matrix decomposition for $F$:
\begin{equation}
\label{fisher_red}
F = \widetilde{A}^\mathsf{T}\widetilde{\Lambda} \left(\widetilde{P}^{-1} + Q^{-1}\mathrm{I\!I}\right)\widetilde{A},
\end{equation}
where $\widetilde{A}$ is essentially the same as $A$, except that the last POVM element $\bm{a}_{n_\gamma \gamma}^\mathsf{T}$ has been removed for each configuration. The matrix $\widetilde{A}$ is therefore of size $\widetilde{n}_\mathrm{tot}\times (N^2-1)$, where $\widetilde{n}_\mathrm{tot} = \sum_\gamma \widetilde{n}_\gamma = n_\mathrm{tot}-M$ is the total number of independent measurement outcomes. Similarly, the $\widetilde{n}_\mathrm{tot}\times\widetilde{n}_\mathrm{tot}$ diagonal matrix $\widetilde{P} = \textrm{diag}\left(\widetilde{\bm{p}}\right)$ is formed from the vector $\widetilde{\bm{p}}$ of probabilities remaining after the last probability $p_{n_\gamma \gamma}$ for each configuration has been removed from $\bm{p}$. The $\widetilde{n}_\mathrm{tot}\times\widetilde{n}_\mathrm{tot}$ diagonal matrix $\widetilde{\Lambda}$ is defined accordingly as
\begin{equation}
\label{LamDef2}
\widetilde{\Lambda} = \textrm{diag}\left(\left[\lambda_1 \mathbb{1}_{\widetilde{n}_1}^\mathsf{T},\ldots,\lambda_\gamma \mathbb{1}_{\widetilde{n}_\gamma}^\mathsf{T},\ldots,\lambda_M \mathbb{1}_{\widetilde{n}_M}^\mathsf{T}\right]^\mathsf{T}\right).
\end{equation}
$Q$ is the $\widetilde{n}_\mathrm{tot}\times \widetilde{n}_\mathrm{tot}$ `complementary' matrix to $\widetilde{P}$, which contains the probabilities we removed from $\widetilde{P}$,
\begin{equation}
\label{Qdef}
Q = \textrm{diag}\left(\left[p_{n_1 1} \mathbb{1}_{\widetilde{n}_1}^\mathsf{T},\ldots,p_{n_\gamma \gamma} \mathbb{1}_{\widetilde{n}_\gamma}^\mathsf{T},\ldots,p_{n_M M} \mathbb{1}_{\widetilde{n}_M}^\mathsf{T}\right]^\mathsf{T}\right).
\end{equation}
Finally, the matrix $\mathrm{I\!I}$ appearing in Eq.~(\ref{fisher_red}) is the $\widetilde{n}_\mathrm{tot}\times\widetilde{n}_\mathrm{tot}$ block-diagonal matrix whose $\gamma^\mathrm{th}$ sub-block is equal to $\mathbb{1}_{\widetilde{n}_\gamma} \mathbb{1}_{\widetilde{n}_\gamma}^\mathsf{T}$.

If the matrix $\widetilde{A}$ is square, it should now be invertible (unless we are unlucky and it is rank-deficient for some other reason, unrelated to the normalization of the POVM elements). To make $\widetilde{A}$ square, we must have that
\begin{equation}
\label{minimal_cond}
\widetilde{n}_\mathrm{tot}=N^2-1.
\end{equation}
That is, we require the total number of independent measurement outcomes to equal the number of independent real parameters specifying the quantum state. If $\widetilde{n}_\mathrm{tot}<N^2-1$, tomographic inversion is not possible (this is intuitively obvious, but the Fisher information is also singular in this case). If $\widetilde{n}_\mathrm{tot}>N^2-1$, the quantum state is over-determined by our measurements. For this reason, we refer to the situation $\widetilde{n}_\mathrm{tot}=N^2-1$ as `minimal tomography'. There are good reasons why over-determined tomography is advantageous, since this introduces redundancy which suppresses the effects of statistical noise \cite{Burgh:2008xu}. However, when attempting tomography of high-dimensional systems, the number of measurements required can become large, and in this case minimal tomography, involving the fewest number of measurements possible, is appealing. Below we present an exact analytic result for the OED for minimal tomography on arbitrarily large systems.

We start by explicitly inverting $F$ in order to obtain the CRB. Define
\begin{equation}
\label{K_def}
K=\widetilde{A}\widetilde{A}^\mathsf{T}
\end{equation}
as the symmetric matrix whose elements are the scalar products between the POVM elements, $K_{\alpha'\gamma,\beta'\delta}=\bm{a}_{\alpha'\gamma}.\bm{a}_{\beta'\delta}$. Then we can write
\begin{eqnarray}
\nonumber B &=& \textrm{tr}\left\{\widetilde{\Lambda}^{-1}K^{-1}\left(\widetilde{P}^{-1}+Q^{-1}\mathrm{I\!I}\right)^{-1}\right\}\\
\label{CRB1} &=& \textrm{tr}\left\{\widetilde{\Lambda}^{-1}K^{-1}\left(\widetilde{P}-\mathbb{P}\right)\right\},
\end{eqnarray}
where in the second line we have inverted $\widetilde{P}^{-1} + Q^{-1}\mathrm{I\!I}$ block-wise using the rank-$1$ update to a matrix inverse \cite{Petersen:2006gf}, introducing the block-diagonal matrix $\mathbb{P}$ whose $\gamma^\mathrm{th}$ sub-block is equal to $\widetilde{\bm{p}}_\gamma\widetilde{\bm{p}}_\gamma^\mathsf{T}$, where $\widetilde{\bm{p}}_\gamma$ is the $\widetilde{n}_\gamma$-dimensional vector of probabilities for all but the last outcome of the $\gamma^\mathrm{th}$ measurement.

After a little algebra, we can re-write this result in the following way,
\begin{equation}
\label{rewriteB}
B = \sum_{\alpha' \gamma} \frac{b_{\alpha'\gamma}}{\lambda_\gamma},
\end{equation}
where the $b_{\alpha'\gamma}$ are the elements of the $\widetilde{n}_\mathrm{tot}$-dimensional vector
\begin{equation}
\label{bdef}
\bm{b}=\widetilde{\bm{p}}*\left(\bm{d}-D\widetilde{\bm{p}}\right),
\end{equation}
with $D$ an $\widetilde{n}_\mathrm{tot}\times\widetilde{n}_\mathrm{tot}$ block-diagonal matrix whose blocks are given by the diagonal blocks of $K^{-1}$, and with the vector $\bm{d}$ given by
\begin{equation}
\label{d_def}
\bm{d}=\textrm{diag}(D) = \textrm{diag}\left(K^{-1}\right).
\end{equation}
Here the symbol $*$ indicates the Hadamard product, \emph{i.e.} element-wise multiplication.

To find the OED, we differentiate the cost function $J$ in Eq.~(\ref{cost1}). Armed with the expression (\ref{rewriteB}) for $B$, we can do this analytically, and we arrive at the result (omitting an unimportant normalization factor)
\begin{equation}
\label{OED}
\lambda^\textrm{OED}_\gamma = \sqrt{\sum_{\alpha'}b_{\alpha'\gamma}}.
\end{equation}
For the special case of binary measurements (such as `click'/`no-click' photon counters), the formula becomes especially simple, since binary POVMs have only a single independent outcome each. The matrix $D$ becomes purely diagonal, and we can write
\begin{equation}
\label{binary}
\lambda^\textrm{OED:binary}_\gamma = \sqrt{d_\gamma p_\gamma(1-p_\gamma)},
\end{equation}
where we have dropped the redundant index $\alpha'$.
\section{Average OED}
\label{section:average}
So far we have considered the problem of OED when the true state is known --- both the optimization in Section \ref{section:algorithm} and the analytic expressions of the previous section require knowledge of $\bm{r}$ in order that the statistics $\bm{p}$ are fixed. What if we really have no information about the true state? This is surely when tomography is most useful, but we cannot then calculate the OED. In this section we will introduce the \emph{average} OED, which does not require any knowledge of $\bm{r}$. The idea is a simple one: where there is a dependence on $\bm{r}$, we average over the space of all possible states, to account for our complete uncertainty about the true state. We will introduce two approaches. In the first, we average the Fisher information before minimizing the resulting CRB. In the second, we calculate the CRB analytically, and then perform the average. It seems that these two approaches produce essentially the same results, which is perhaps natural (but not obvious!). Using the first approach, one can calculate the average OED using a numerical optimization similar to the one described in Section \ref{section:algorithm}. The second approach yields an analytic result for the average OED, but of course it can only be applied to minimal tomography.

Both approaches, however, are exact \emph{only for qubits}. The reason is that the averaging requires an integral over the space of physical states. This space is the $(N^2-1)$-dimensional volume such that all states $\bm{r}$ within it correspond to positive density matrices. Evaluating the boundaries of this region is non-trivial, and so in general the averaging cannot be performed exactly. But it is known that the space of physical states is a simply-connected convex region, whose boundary lies between two concentric $(N^2-2)$-dimensional hyperspheres whose radii are given respectively by \cite{Harriman:1967kx}
\begin{equation}
\label{Rs}
R_\mathrm{min}=\frac{1}{\sqrt{N(N-1)}} \quad \textrm{and}\quad R_\mathrm{max}=\sqrt{\frac{N-1}{N}}.
\end{equation}
For the case of qubit tomography, with $N=2$, these hyperspheres coincide, with $R_\mathrm{min} = R_\mathrm{max}=R_2 = 1/\sqrt{2}$, the radius of the well-known Bloch sphere on which all pure qubit states lie. For higher dimensional systems, we will continue to approximate the space of physical states as being bounded by a hypersphere of radius $R_N$, where $R_N$ must lie somewhere in between $R_\mathrm{min}$ and $R_\mathrm{max}$. If, at the end of the calculation, we find that the average OED does not depend (or depends only weakly) on our choice for $R_N$, then we can be confident that we have closely approximated the `true' average OED. In practice, we have found this to be the case, so that our averaging procedure --- crude though it is --- produces useful results.

\subsection{Averaging the Fisher information}
\label{subsection:averaging_fish}
The definition of the Fisher information in Eq.~(\ref{fisher}) involves the evaluation of an ensemble average. For calculating the OED, we assumed that every member of our ensemble was prepared in the same state, and so we averaged over the statistical distribution of the data $\bm{n}$, assuming a given state $\bm{r}$. If we consider that, in fact, the prepared state is itself drawn from a large ensemble of possibilities, then we should extend our calculation of the expectation by averaging our result over those possibilities. With no prior information at all, we should average over the space of all physical states. The Fisher information obtained by averaging in this way is
\begin{equation}
\label{average_fish}
\langle F\rangle = A^\mathsf{T}\Lambda G A,
\end{equation}
where the elements of the diagonal matrix $G=\langle P^{-1}\rangle$ are given by
\begin{equation}
\label{Pelements}
g_{\alpha\gamma}=\int \frac{\mathfrak{p}(\bm{r})}{p_{\alpha\gamma}(\bm{r})}\,\mathrm{d}V,
\end{equation}
with $\mathfrak{p}$ the probability distribution from which the states are drawn. In what follows we consider a uniform distribution $\mathfrak{p}=\textrm{const}$, and take the integral to run over the $(N^2-1)$-dimensional volume enclosed by a hypersphere of radius $R_N$ centred at the origin. For $N=2$ the integral in Eq.~(\ref{Pelements}) can be evaluated analytically to give
\begin{eqnarray}
\nonumber
g_{\alpha \gamma} &=& \frac{3}{4R_2^3|\bm{a}_{\alpha\gamma}|^3}\bigg\{ \left(c_{\alpha\gamma}^2 - R_2^2|\bm{a}_{\alpha\gamma}|^2\right)\ln \left[\frac{c_{\alpha\gamma}-R_2|\bm{a}_{\alpha\gamma}|}{c_{\alpha\gamma}+R_2|\bm{a}_{\alpha\gamma}|}\right]\\
\label{bloch_int}& &+2c_{\alpha\gamma}R_2|\bm{a}_{\alpha\gamma}|\bigg\}.
\end{eqnarray}
The generalization to higher dimensions can be found recursively,
\begin{equation}
\label{higher_int}
g_{\alpha\gamma}=\frac{1}{|\bm{a}_{\alpha\gamma}|}\times\left\{ I_{N^2-3}\left(c_{\alpha\gamma},|\bm{a}_{\alpha\gamma}|\right)-I_{N^2-3}\left(c_{\alpha\gamma},-|\bm{a}_{\alpha\gamma}|\right)\right\},
\end{equation}
where the integral $I_n(a,b)=\int_0^{R_N} x^n\ln\left(a+bx\right)\,\mathrm{d}x$ satisfies the recurrence relation
\begin{eqnarray}
\nonumber
(n+1)I_n(a,b) &=& \frac{R_N^{n}}{b}\left\{(a+bR_N)\left[\ln\left(a+bR_N\right)-1\right] + a\right\} \\
\label{recurrence} & &+ R_N^{n+1}\frac{n}{n+1}-n\frac{a}{b}I_{n-1}(a,b),
\end{eqnarray}
with $I_0(a,b) = \frac{1}{b}(a+bR_N)\left[\ln\left(a+bR_N\right)-1\right]-\frac{a}{b}\left[\ln(a)-1\right]$. Once the elements $g_\ag$ of $G$ are known, we can apply the numerical method described in Section \ref{section:algorithm} to find the average OED, which is the experiment design $\bm{\lambda}^{\langle\mathrm{OED}\rangle}$ that minimizes the associated CRB $\langle B\rangle = \textrm{tr}\left(\langle F \rangle^{-1}\right)$. The only difference is that the matrix $P^{-1}$ in Eqs.~(\ref{DiffJ}), (\ref{hessian}) and (\ref{Kdef}) should be replaced with $G$.

In Figure \ref{figure:numerical_average_OED} below, we show the result of such a numerical optimization, for the same set of POVMs as were used to generate Figure \ref{figure:compare_Kosut} in Section \ref{section:algorithm} (the model is described in \cite{Kosut:2004pi}). It is notable that the two figures look very different: the OED of Figure \ref{figure:compare_Kosut} is optimal for a particular state (a pure state in this case), while the average OED has been constructed so as to be optimal for any state, or more precisely, optimal when we do not know which state has been prepared.
\begin{figure}[h]
\begin{center}
\includegraphics[width=\columnwidth]{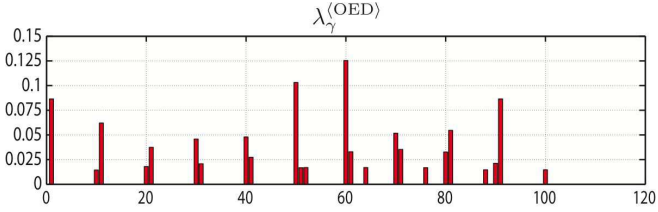}
\caption{The average OED computed numerically from the averaged Fisher information. The set of POVMs is the same as those used to generate Figure \ref{figure:compare_Kosut}. The marked difference between the two experiment designs shows how ignorance of the true state changes the optimization.}
\label{figure:numerical_average_OED}
\end{center}
\end{figure}

If we are interested in minimal tomography (so that Eq.~(\ref{minimal_cond}) is satisfied), we can write down an analytic formula for the average OED. The averaged Fisher information is
\begin{equation}
\label{average_fisher_analytic}
\langle F\rangle = \widetilde{A}^\mathsf{T}\widetilde{\Lambda} \left\langle \widetilde{P}^{-1} + Q^{-1}\mathrm{I\!I}\right\rangle \widetilde{A}.
\end{equation}
Inverting this gives the associated CRB,
\begin{equation}
\label{av_B}
\langle B\rangle = \sum_{\alpha'\gamma} \frac{\langle b_{\alpha'\gamma}\rangle}{\lambda_\gamma},
\end{equation}
where the averaging procedure yields
\begin{equation}
\label{av_b_def}
\langle \bm{b}\rangle = \widetilde{\bm{g}}^{-1}*\left(\bm{d}-\bm{f}*D\widetilde{\bm{g}}^{-1}\right),
\end{equation}
with $\widetilde{\bm{g}}^{-1}$ the element-wise inverse of the vector $\widetilde{\bm{g}}$, whose elements are the $g_{\alpha'\gamma}$, defined in Eqs.~(\ref{bloch_int}) and (\ref{higher_int}) above. The tilde symbol indicates, as usual, that the element associated with the last outcome of each POVM has been removed. The $\widetilde{n}_\mathrm{tot}$-dimensional vector $\bm{f}$ is defined as follows,
\begin{equation}
\label{f_def}
f_{\alpha'\gamma} = \frac{1}{\sum_{\beta} g_{\beta \gamma}^{-1}},\quad\textrm{for all}\; \alpha'.
\end{equation}

Differentiation of the appropriate cost function then yields the formula for the average OED (again omitting a normalization factor),
\begin{equation}
\label{avOEDform}
\lambda^{\langle \mathrm{OED}\rangle}_\gamma = \sqrt{\sum_{\alpha'}\langle b_{\alpha'\gamma}\rangle}.
\end{equation}
For the case of binary measurements, we drop the index $\alpha'$, and we have $f_\gamma = g_\gamma$. The formula in Eq.~(\ref{av_b_def}) then vanishes,
$$
\langle b_\gamma \rangle = \frac{d_\gamma}{g_\gamma}\left(1-\frac{g_\gamma}{g_\gamma}\right) = 0,
$$
but the term in brackets cancels when the $\lambda_\gamma$ are normalized, so we arrive at the simple expression
\begin{equation}
\label{avOEDform_binary}
\lambda^{\langle \mathrm{OED:binary}\rangle}_\gamma = \sqrt{d_\gamma/g_\gamma}.
\end{equation}

\subsection{Averaging the CRB directly}
We now describe a second method for computing the average OED. Our approach here is to average the analytic expression in Eq.~(\ref{rewriteB}) for the CRB over a hypersphere --- our approximation to the space of all physical states. Since the analytic expression only holds for minimal tomography, when Eq.~(\ref{minimal_cond}) is satisfied, this method is applicable only in this situation. To distinguish this from the procedure outlined above, where we averaged $F$, we will use two angle-brackets to denote this type of average:
\begin{equation}
\label{direct_av}
\dubav{B} = \sum_{\alpha' \gamma} \frac{\dubav{ b_{\alpha'\gamma} } }{\lambda_\gamma}.
\end{equation}
Performing this averaging yields the result
\begin{equation}
\label{direct_av_result}
\dubav{\bm{b}} = \bm{c}*\left(\bm{d}-D\bm{c}\right) - \dubav{x^2} \textrm{diag}\left(DK\right),
\end{equation}
where $\dubav{x^2}$ stands for the integral of a Cartesian coordinate $x^2$ over the hypersphere, divided by the hypersphere volume. For qubits, with $N=2$, we have $\dubav{x^2} = 1/10$, and in general we find \cite{Folland:2001jt,Abramowitz:1965rp}
\begin{equation}
\label{x2av}
\dubav{x^2}=\frac{R_N^2 (N^2-1)\Gamma\left(\tfrac{N^2-1}{2}\right)}{2(N^2+1)\Gamma\left(\tfrac{N^2+1}{2}\right)},
\end{equation}
where $\Gamma(.)$ is the Euler Gamma function. Differentiating the appropriate cost function leads to the following analytic formula for the average OED,
\begin{equation}
\label{direct_av_lam}
\lambda^{\dubav{\textrm{OED}}}_\gamma = \sqrt{\sum_{\alpha'} \dubav{b_{\alpha'\gamma}}}.
\end{equation}
For the case of binary measurements, the formula reduces to
\begin{equation}
\label{direct_av_lam_binary}
\lambda^{\dubav{\textrm{OED:binary}}}_\gamma = \sqrt{d_\gamma \left[c_\gamma(1-c_\gamma)-\dubav{x^2}|\bm{a}_\gamma|^2\right]}.
\end{equation}

As a simple application of the these approaches to calculating the average OED, consider the particular case of binary measurements chosen so that the POVM elements are mutually orthogonal in the Bloch representation, with $\bm{a}_\gamma.\bm{a}_\delta = a^2\delta_{\gamma\delta}$, for some constant $a$. Then the matrix $K$ of scalar products is proportional to the identity matrix, and inverting it gives $d_\gamma = 1/a^2$ for all measurements. Suppose in addition that the measurements are symmetric in the sense that the `identity components' associated with the two outcomes of each measurement are the same, $c_\gamma = 1-c_\gamma =1/2$. Then Eqs.~(\ref{direct_av_lam_binary}) and (\ref{avOEDform_binary}) give
\begin{eqnarray}
\nonumber \lambda^{\dubav{\textrm{OED:binary}}}_\gamma &=& \sqrt{\frac{1}{4a^2}-\dubav{x^2}}\\
\nonumber &=& \textrm{constant},\\
\nonumber \textrm{and}\quad \lambda^{\langle \mathrm{OED:binary}\rangle}_\gamma &=& \frac{1}{\sqrt{a\left[I_{N^2-3}(\frac{1}{2},a)-I_{N^2-3}(\frac{1}{2},-a)\right]}} \\
\label{constant} &=& \textrm{constant}.
\end{eqnarray}
That is, both approaches to computing the average OED predict a uniform distribution. This is to be expected, because the measurements just described are the most symmetric binary measurements possible. For $N=2$, they constitute projective measurements chosen from the three mutually unbiased bases (MUBs \cite{Paterek:2002om}) for a qubit. It is already known that these MUB measurements are optimal in the sense that, given a uniform distribution for the experiment design, and no prior knowledge of the state, they provide the best tomographic precision \cite{Wootters:1989cj}. The above result provides an alternative perspective: MUB measurements are those for which the best tomographic precision, without prior knowledge of the state, is achieved using a uniform experiment design.

\begin{figure}[h]
\begin{center}
\includegraphics[width=\columnwidth]{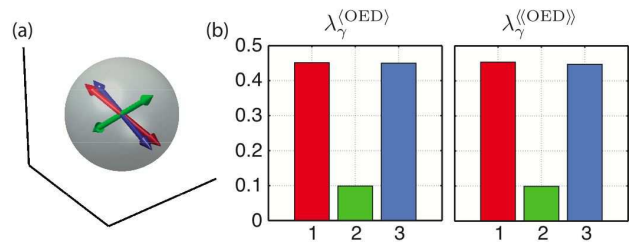}
\caption{Comparison of the two methods for calculating the average OED. (a): the set of 3 binary POVMs used in the qubit tomography set-up. Arrows represent the vectors $\bm{a}_\ag$ on the Bloch sphere. (b): On the left is the average OED predicted by averaging the Fisher information (see Eq.~(\ref{avOEDform})); on the right is the corresponding result predicted by averaging the CRB (see Eq.~(\ref{direct_av_lam})). The two results are indistinguishable. This is typical: sometimes differences between the two methods are discernible, but generally they produce very similar experiment designs.}
\label{figure:compare_averages}
\end{center}
\end{figure}

In Figure \ref{figure:compare_averages}, we plot the average OED $\bm{\lambda}^{\dubav{\textrm{OED}}}$ predicted by Eq.~(\ref{direct_av_lam}), alongside the prediction $\bm{\lambda}^{\langle \mathrm{OED}\rangle}$ of the method presented previously (Eq.~(\ref{avOEDform})), for the case of minimal qubit tomography using a randomly chosen set of $3$ binary POVMs. Note that these results are exact, since the state space is exactly spherical for a qubit. What is notable is that the two predictions appear to coincide --- although the formulae are different, the two averaging procedures seem to result in identical, or very similar, experiment designs. Conceptually, minimizing the averaged CRB is slightly different to minimizing the CRB associated to the averaged Fisher information, so it is not immediately obvious why this is so. Empirically, however, we have found that both methods generate the same results.

\section{ODT}
\label{section:odt}
The method just presented has an obvious generalization: instead of choosing $\bm{\lambda}$ so as to minimize the averaged Cramer-Rao bound $\dubav{B}$, we could find an expression for $\dubav{\delta B^2}$, the \emph{variance} of $B$ over the state space, and try to find the vector $\bm{\lambda}$ that minimizes this variance. Such an experiment design makes the tomographic reconstruction as `fair' as possible, since the reconstruction should have as close as possible to a uniform precision for all states. We call this experiment design the \emph{optimal design for tomography}, or ODT for short. Using the expression in Eq.~(\ref{rewriteB}), we can write the variance of $B$ as
\begin{eqnarray}
\nonumber \dubav{\delta B^2} &=& \dubav{B^2}-\dubav{B}^2\\
\label{var1} &=& \bm{\lambda}^{-1\mathsf{T}}V\bm{\lambda}^{-1},
\end{eqnarray}
where $\bm{\lambda}^{-1}$ is the element-wise inverse of $\bm{\lambda}$, and where the real, symmetric (and therefore positive definite) $M\times M$ variance matrix $V$ is given by
\begin{eqnarray}
\nonumber
V_{\gamma \delta} &=& \sum_{\alpha'\beta'}\dubav{b_{\alpha'\gamma}b_{\beta'\delta}}-\dubav{b_{\alpha'\gamma}}\dubav{b_{\beta'\delta}}\\
\label{var_mat} &=& \sum_{\alpha'\beta'}v_{\alpha'\gamma,\beta'\delta}.
\end{eqnarray}
Performing the average explicitly requires a stout heart; the result is
\begin{eqnarray}
\nonumber v_{\alpha'\gamma,\beta'\delta} &=& \dubav{x^2}d_{\alpha'\gamma}d_{\beta'\delta}K_{\alpha'\gamma,\beta'\delta} \\
\nonumber & & -2\dubav{x^2}\sum_j c_{\alpha'\gamma}d_{\beta'\delta}W_{\alpha'\gamma,j}\widetilde{A}_{\beta'\delta,j} \\
\nonumber & & +4\dubav{x^2}\sum_j c_{\alpha'\gamma}c_{\beta'\delta}W_{\alpha'\gamma,j}W_{\beta'\delta,j}\\
\nonumber & & +\dubav{x^4}\sum_j X_{\alpha'\gamma,j} X_{\beta'\delta,j}\\
\nonumber & & +\dubav{x^2y^2}\sum_{j\neq k} X_{\alpha'\gamma,j}X_{\beta'\delta,k}\\
\nonumber & & + 2\dubav{x^2y^2}\sum_{j\neq k} \widetilde{A}_{\alpha'\gamma,j} \widetilde{A}_{\beta'\delta,j} W_{\alpha'\gamma,k} W_{\beta'\delta,k}\\
\label{big_var}& & -\dubav{x^2}\sum_{j\neq k} X_{\alpha'\gamma,j}X_{\beta'\delta,k},
\end{eqnarray}
where we have defined the matrices $W=D\widetilde{A}$ and $X=\widetilde{A}*W$, and where the roman indices $j,k$ number the Cartesian coordinates of the state space, and run from $1$ to $N^2-1$. The average $\dubav{x^2}$ is defined in Eq.~(\ref{x2av}) and the other averages are given by \cite{Folland:2001jt,Abramowitz:1965rp}
\begin{eqnarray}
\nonumber \dubav{x^2y^2}&=&\frac{R_N^4(N^2-1)\Gamma\left(\tfrac{N^2-1}{2}\right)}{4(N^2+3)\Gamma\left(\tfrac{N^2+3}{2}\right)},\\
\label{other_avs}\dubav{x^4}&=&3\dubav{x^2y^2}.
\end{eqnarray}
We can now find the experiment design that minimizes Eq.~(\ref{var1}). Setting the derivative of the cost function $J=\dubav{\delta B^2}+\eta\left(\sum_\gamma \lambda_\gamma -1\right)$ to zero, we obtain the condition
\begin{equation}
\label{var_cond}
V\bm{\lambda}^{-1}=\eta \bm{\lambda}^2,
\end{equation}
where $\bm{\lambda}^2$ is a column vector whose elements are given by $\lambda_\gamma^2$. It is not clear how to solve Eq.~(\ref{var_cond}) analytically, but it is not hard to find the solution $\bm{\lambda}^{\mathrm{ODT}}$ numerically --- almost any optimization algorithm will suffice.

\subsection{ODT or average OED?}
Both ODT and average OED are reasonable candidates for an experiment design that does not rely on prior knowledge of the true state $\rho$. They are both the results of a minimization, but their objective functions are different, so how do they compare? To contrast the two optimizations, we used the analytic results in Eqs.~(\ref{direct_av}) and (\ref{var1}) to evaluate the objective functions $\dubav{B}$ and $\dubav{\delta B^2}$ using many sets of $N^2-1$ randomly chosen binary POVMs, for a range of dimensionalities $N=2$--$10$. Operationally, the quantities $\sqrt{\dubav{B}}$ and $\dubav{\delta B^2}^{1/4}$ are more intuitive, having natural interpretations as distances in the Bloch representation. The former is the bound on the root-mean-squared (r.m.s.) error in the reconstructed state $\hat{\bm{r}}$ --- the typical value of $\left|\hat{\bm{r}}-\bm{r}\right|$; the latter is the square root of the standard deviation of the CRB, which can be thought of as the range over which the r.m.s. error $\left|\hat{\bm{r}}-\bm{r}\right|$ varies.

In Figure \ref{figure:performance} below, we show in the top two panels how the average OED and the ODT improve (\emph{i.e.} reduce) the averages of these quantities (averaged over $3000$ sets of random POVMs) when compared against the results generated with a uniform experiment design $\bm{\lambda} = \mathbb{1}_M/M$. The improvements are all at the level of around $10\%$, for both the r.m.s. error $\sqrt{\dubav{B}}$ when using $\bm{\lambda}^{\dubav{\mathrm{OED}}}$ (part (a)), and the `deviation' $\dubav{\delta B^2}^{1/4}$ when using $\bm{\lambda}^{\mathrm{ODT}}$ (part (b)). Improvements at this level are certainly significant enough to motivate spending the time to calculate the appropriate optimal design.

On the other hand, the lower two panels show how the two designs $\bm{\lambda}^{\dubav{\mathrm{OED}}}$ and $\bm{\lambda}^{\mathrm{ODT}}$ perform in terms of the other's objective function. Part (c) shows that the average OED only improves the r.m.s. error by around $1\%$ over that achieved by the ODT. Similarly part (d) shows that the ODT, on average, reduces the deviation $\dubav{\delta B^2}^{1/4}$ by around $1\%$ with respect to the deviation produced by using the average OED. And the differences fall away as the dimension $N$ increases. It seems that there is actually not much to choose between the ODT and the average OED in terms of performance. This is good news: although there \emph{is} a tradeoff between precision and fairness, it is small in the sense that optimizing for one only reduces the other by around a percent or less.

These results are not definitive, because we cannot integrate over the true state space --- only our hyperspherical approximations to it --- but it is clear from Figure~\ref{figure:performance} that the choice made for $R_N$ has little influence on our conclusion, suggesting that it should hold for the correctly averaged quantities too.

\begin{figure}[h]
\begin{center}
\includegraphics[width = \columnwidth]{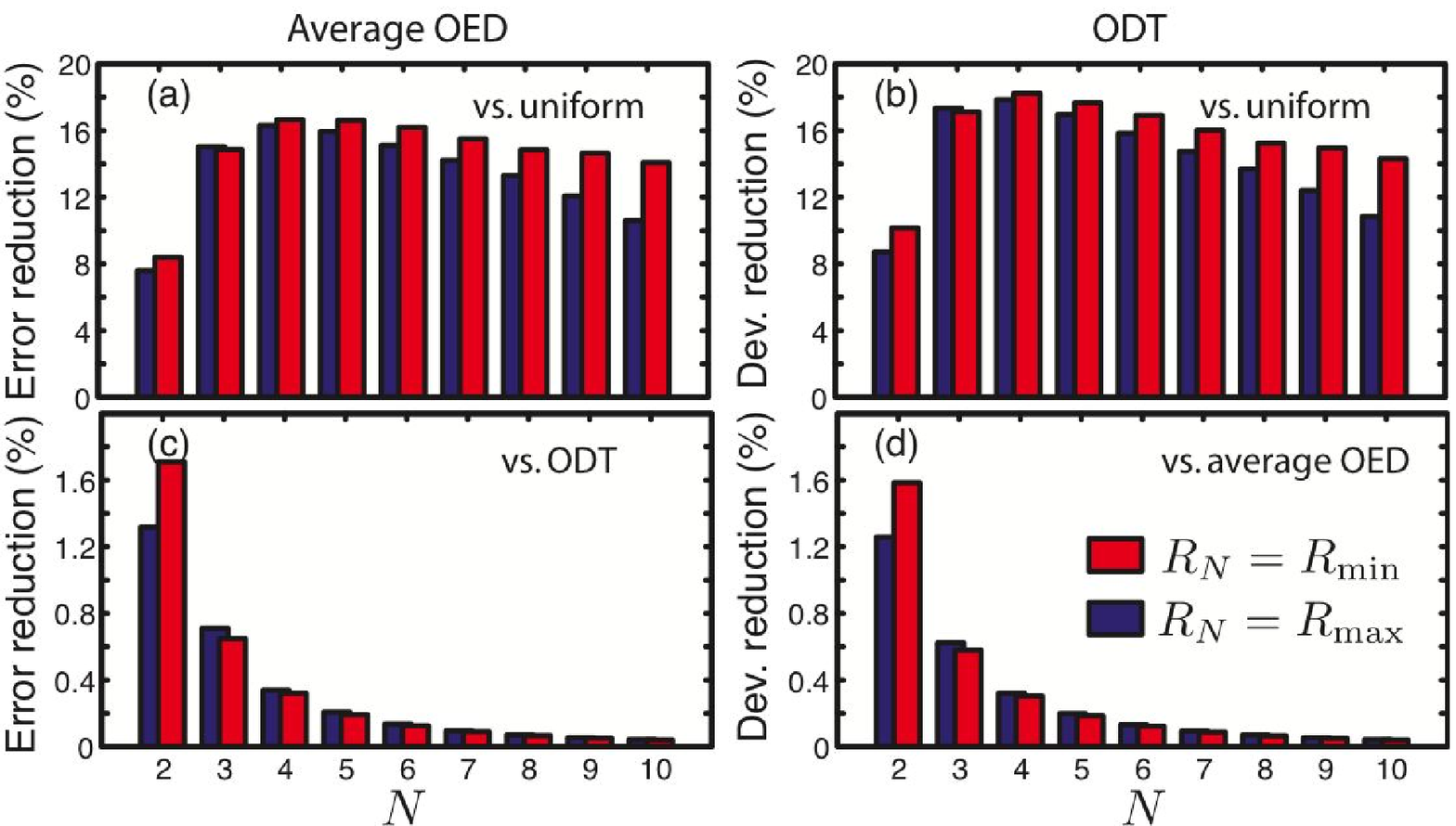}
\caption{Comparison of ODT and average OED. The top two panels show how (a) average OED and (b) ODT compare with a uniform design $\lambda_\gamma = 1/M$. In part (a), the relative reduction in the average of the r.m.s. error $\sqrt{\dubav{B}}$ afforded by using $\bm{\lambda}^{\dubav{\mathrm{OED}}}$ instead of a uniform design, averaged over $3000$ sets of randomly generated POVMs, is plotted against the dimension $N$ of the quantum systems: the red bars result from setting $R_N = R_\mathrm{min}$; the blue from setting $R_N = R_\mathrm{max}$. In part (b), we show the relative percentage reduction in the average of the `deviation' $\dubav{\delta B^2}^{1/4}$ afforded by using $\bm{\lambda}^{\mathrm{ODT}}$ instead of a uniform design. The lower two plots compare ODT and average OED with each other. In part (c), we plot the relative reduction in the average r.m.s. error arising from using the average OED instead of the ODT. In part (d), the relative reduction in the deviation given by choosing the ODT over the average OED is shown. Note that the averages $\dubav{.}$ over the state space are exact for qubits, for which $R_\mathrm{min}=R_\mathrm{max}$, so the discrepancy between the red and blue bars is entirely statistical for $N=2$; the results for higher dimensions should be interpreted with this in mind.}
\label{figure:performance}
\end{center}
\end{figure}

\section{Monte Carlo simulations}
\label{section:monte_carlo}
So far we have presented a number of mathematical results involving approximate averages of asymptotic bounds. The reader could be forgiven for doubting the utility of these results in the laboratory. Short of performing real experiments --- the ideal proving ground --- we cannot do better than applying our techniques to simulated data, so we have performed some Monte-Carlo simulations of minimal qubit tomography that provide some insight into when the CRB is saturated, and how much is gained by implementing an optimized experiment design.

\begin{figure}[h]
\begin{center}
\includegraphics[width=6cm]{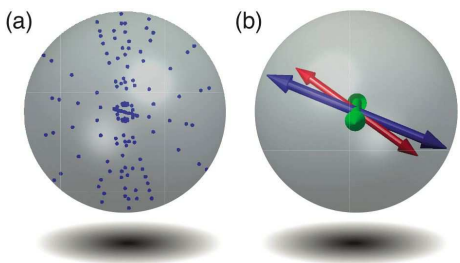}
\caption{(a) The 216 sample states used to construct averages over the Bloch sphere using polynomial interpolation. The points are distributed evenly around the equator, but Chebyshev grids are used for the polar and radial distributions --- this explains the clustering of points towards the poles and the centre of the Bloch sphere. (b) The POVM elements used in the Monte-Carlo simulations presented below.}
\label{figure:states}
\end{center}
\end{figure}

In these simulations, we generate a set of qubit states distributed throughout the Bloch sphere, as shown in part (a) of Figure \ref{figure:states}. The states are chosen so that it is easy to calculate averages over the Bloch sphere by polynomial interpolation: we use Chebyshev grids for the radial and polar coordinates and an equally spaced grid for the azimuth; averages over the sphere can then be computed accurately using Clenshaw-Curtis quadrature radially and polar-wise, and Fourier interpolation around the equator \cite{Trefethen:2000oq}. Using $6$ points for each coordinate yields averages that are accurate to within about $1\%$ with reasonable computing times.

\begin{figure}[h]
\begin{center}
\includegraphics[width=\columnwidth]{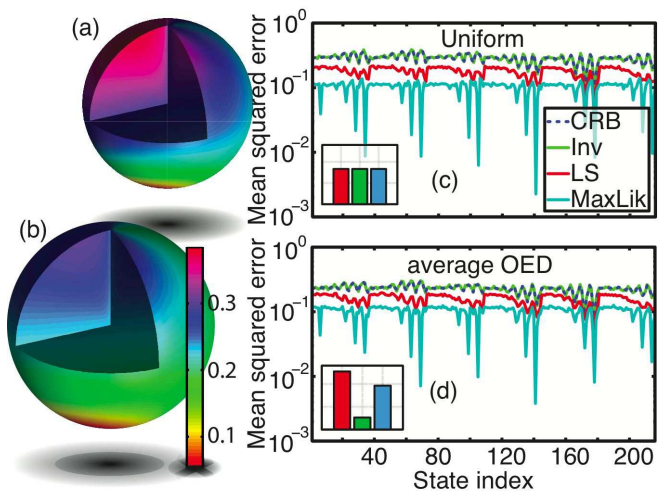}
\caption{Monte-Carlo results. Parts (a) and (b) show the variation of the CRB over the Bloch sphere for the uniform design and the average OED, respectively. In part (c), the mean squared errors achieved by the three methods inversion, least squares and maximum likelihood are plotted for the $216$ sample states, along with the CRB (dashed line), for a uniform experiment design (inset). Part (d) shows the same results, except that the average OED was used instead (inset). The plots have an oscillatory appearance because they are produced by `unwrapping' the Bloch sphere, with the sample states from part (a) of figure (\ref{figure:states}) arranged side by side.}
\label{figure:monte_carlo_compare}
\end{center}
\end{figure}

Figure \ref{figure:monte_carlo_compare} shows the mean squared errors predicted by the CRB alongside the errors actually achieved using simulated data equivalent to $2000$ experimental runs, each consisting of $N_\mathrm{tot}=1000$ measurements. We chose a minimal set of binary POVMs randomly (shown in part (b) of Figure \ref{figure:states}), and then performed two simulations --- one with a uniform experiment design, and one using the average OED predicted by Eq.~(\ref{direct_av_lam}). Results are plotted for three different methods of state reconstruction, which we call \emph{inversion}, \emph{least squares} and \emph{maximum likelihood} \cite{Kaznady:2009cl,Hradil:2006ud}.

\subsection{Reconstruction methods}
\label{subsection:reconstruction_methods}
Inversion is the simplest: we first generate an estimate $\overline{\bm{p}}$ of the `true' statistics $\bm{p}$ using the relative frequencies of each measurement outcome in the simulated data $\{n_\ag\}$, by setting $\overline{p}_\ag=n_\ag/N_\gamma$. We then substitute this into Eq.~(\ref{prob3}) and solve for the Bloch vector,
\begin{equation}
\label{inversion}
\hat{\bm{r}} = A^{-}\left(\overline{\bm{p}}-\bm{c}\right).
\end{equation}
Here the notation $A^{-}$ stands for the Moore-Penrose pseudo-inverse of $A$ \cite{Albert:ys,Golub:1965rr,Greville:1960fr}, which exists even though $A$ is generally rectangular (and therefore has no true inverse). In Matlab we use the backslash operator \cite{Recktenwald:2000jt}, which computes $\hat{\bm{r}}$ directly by Gaussian elimination. A problem with direct inversion is that sometimes the estimate $\hat{\bm{r}}$ lies outside the Bloch sphere, producing an unphysical reconstructed density matrix. In the least squares method, we remove any of these unphysical estimates by using the `nearest' (in the least squares sense) physical state. This is particularly easy for the qubit states we consider here, since the state space is spherical: whenever the norm of $\hat{\bm{r}}$ exceeds $R_2=1/\sqrt{2}$, we re-normalize it,
\begin{equation}
\label{least_squares}
\hat{\bm{r}}\longrightarrow R_2\times \frac{\hat{\bm{r}}}{\left|\hat{\bm{r}}\right|}.
\end{equation}
The maximum likelihood method is a more nuanced approach to the same problem \cite{Banaszek:1999la,Hradil:2004gd,Paris:2004kx,Kosut:2004pi,Lvovsky:2004it}. The rationale is to try to find the physical state which is most likely to have produced the observed (simulated) data. This is the state $\hat{\bm{r}}$ that maximizes the likelihood function $p\left(\bm{n}|\hat{\bm{r}}\right)$ defined in Eq.~(\ref{prob_data}), or --- more conveniently --- the state that renders the corresponding log-likelihood $\hat{\mathcal{L}}$ (see Eq.~(\ref{loglik})) stationary with respect to changes in the estimated density matrix $\hat{\rho}$. A simple iterative scheme that converges on this state, while including the positivity and trace constraints on $\hat{\rho}$, has been derived by Hradil. Starting with an unphysical state $\hat{\bm{r}}$ generated by inversion, we first normalize it as per the least squares method. We then construct the corresponding density matrix $\hat{\rho}$ using Eq.~(\ref{rho1}), and then make the replacement
\begin{equation}
\label{Hradil}
\hat{\rho}\longrightarrow \frac{1}{2}\left[R\left(\hat{\rho}\right)\hat{\rho} + \hat{\rho}R\left(\hat{\rho}\right)\right],
\end{equation}
where the matrix $R$, which depends on $\hat{\rho}$ through the probabilities $\hat{p}_\ag =  \textrm{tr}\left\{\Pi_\ag \hat{\rho}\right\}$, is given by \cite{Hradil:2004gd,Paris:2004kx}
\begin{equation}
\label{Rdef}
R = \frac{1}{M}\sum_\ag \frac{\overline{p}_\ag}{\hat{p}_\ag}\Pi_\ag.
\end{equation}
We repeat this procedure with the updated estimate, and its associated operator $R$, until the algorithm converges. The resulting state is guaranteed to be physical, where the initial state was not. In the cases that inversion produces a physical estimate to start with, we accept it without applying the above procedure, since it can be shown that under these circumstances inversion already produces the most likely estimate \cite{Paris:2004kx}.
\subsection{Results}
\label{subsection:results}
As is clear from the plots in parts (c) and (d) in Figure \ref{figure:monte_carlo_compare}, the CRB describes the errors achieved by inversion extremely well (in both plots, the two lines are nearly indistinguishable, save for statistical fluctuations). Comparing the two plots, it is also clear that the average OED reduces the inversion errors significantly over those achieved using a uniform design. Integrating over the Bloch sphere, we find for this example that the average r.m.s. error is reduced by $\sim 11\%$ by optimizing the design, in agreement with the predictions shown in Figure~\ref{figure:performance} in the previous section.

However, the precision of the other two reconstruction methods is not well described by the CRB. The least squares method performs significantly better than inversion, while the maximum likelihood method leaves both other methods standing, achieving mean squared errors nearly two orders of magnitude smaller than inversion for some states. The reason why these methods perform better is clear: knowledge of the boundaries of the space of physical states is utilized to improve the tomographic reconstruction. The least squares method is too crude to take full advantage of this constraint, but the maximum likelihood method exploits it to impressive effect. The positivity constraint on the density matrix is not present anywhere in our derivation of the CRB, and this is why an estimation method which implements this constraint is able to beat the `lower bound' on the average errors represented by the CRB. These simulations highlight the fact that the CRB as presented in this paper applies rigorously only to state tomography via direct linear inversion of the measured statistics.

Does this limit the applicability of our results relating to optimal experiment design? In the above example, use of the average OED only improves the average r.m.s. error achieved by the least squares method over a uniform design by around $7\%$, while the maximum likelihood errors are hardly affected at all --- the improvement is less than $1\%$. In the next section, we will consider an alternative formulation of the CRB that includes the positivity constraint, and which therefore correctly describes the precision achieved by the maximum likelihood method. We will see that its optimization results in very similar experiment designs to those presented above, from which we conclude that our results remain close to optimal even when used with estimation methods that enforce positivity.

\section{Constrained estimators}
\label{section:cholesky}
It is well-known in classical statistics that the bias and variance of an estimator are complementary. The precision of an estimate can be improved using some prior knowledge of the estimated quantities, but inevitably this biases the estimate, shifting the mean of the estimator away from the `true' value \cite{Hero-III:1996kl,Fessler:1996cl}. The least squares and maximum likelihood methods are biased estimation techniques, because they incorporate the positivity constraint that excludes unphysical states, but at the same time they are more precise than direct inversion --- which is unbiased --- and they are able to beat the unbiased CRB.

It is useful to visualize the true quantum state as a point in the Bloch sphere, surrounded by a spherical `bubble' that represents isotropic statistical fluctuations. If the true state lies close to the boundary of the Bloch sphere, the bubble may extend into the region of unphysical states. A constrained estimation method such as maximum likelihood will exclude these unphysical states, which distorts the shape of the error bubble. This reduces its size, improving the tomographic precision, but it is clear that it also introduces bias, since the centre of mass of the bubble no longer coincides with the true state.

The CRB can be modified to yield a bound on the precision of biased estimators, but it requires an analytic expression for the gradient of the bias as a function of the true state \cite{Fessler:1993rw,Matson:2006vf}. This depends on the particular estimator being used, which removes some of the generality of the expression; in the case of the maximum likelihood method, it is not possible to write down such an expression in any case, since the estimator itself is only defined implicitly, as the fixed point of the iteration described in Eq.~(\ref{Hradil}) \cite{Fessler:1996cl}. In general, constructing CRBs for estimators biased by inequality constraints --- such as positivity --- is a hard problem \cite{Matson:2006vf}, and so it is not obvious how to adapt the foregoing analysis to maximum likelihood tomography.

In this section, we show that an alternative parameterization of the density matrix allows us to replace the inequality constraint of positivity with the \emph{equality} constraint of unit trace. The CRB for estimators with equality constraints is a much more tractable problem, and we will derive the appropriate constrained CRB (CCRB), which correctly describes maximum likelihood tomography.

\subsection{The Cholesky representation}
\label{subsection:cholesky}
Instead of using the Bloch representation of $\rho$, we will use its \emph{Cholesky decomposition} \cite{Trefethen:1997sf,Banaszek:1999la,Kaznady:2009cl}. That is, we write $\rho$ in the form
\begin{equation}
\label{cholesky}
\rho = T^\dagger T,
\end{equation}
where $T$ is a unique \footnote{$T$ is only unique when $\rho$ is of full rank, which condition excludes pure states, but in such cases one can recover the uniqueness of $T$ by adopting a consistent convention on the positions of zeros along its main diagonal.} upper triangular matrix with real elements along its main diagonal. The positivity of $\rho$ is built-in to this parameterization, because the eigenvalues of $\rho$ are given by the squares of the singular values of $T$. Clearly the Hermiticity of $\rho$ is also guaranteed. However the trace condition must be added separately (compare this with the Bloch representation, in which the trace condition and Hermiticity are `automatic', and positivity must be imposed separately),
\begin{equation}
\label{trace_cond}
\textrm{tr}\left\{T^\dagger T\right\}=1.
\end{equation}
The problem of quantum state tomography now reduces to finding the $N^2$ real numbers that parameterize the matrix $T$, subject to the equality constraint in Eq.~(\ref{trace_cond}). We define the Cholesky vector $\bm{\theta}$ as the column vector comprising these numbers, by analogy with the Bloch vector $\bm{r}$. To construct $\bm{\theta}$, we first define
\begin{equation}
\label{t_def}
\bm{t}=\left[\begin{array}{c}\textbf{vec}\left(\Re\left\{T^\dagger\right\}\right)\\ \textbf{vec}\left(\Im\left\{T^\dagger\right\}\right)\end{array}\right]
\end{equation}
as the $(2N^2)$-dimensional column vector formed by vectorizing first the real part, and then the imaginary part of $T^\dagger$, and concatenating the two. We then eliminate all the redundant elements of $\bm{t}$, which contain zeros because of the structure of $T$ (that is, because $T$ is upper triangular with a real diagonal --- `accidental' zeros we keep). This gives us $\bm{\theta}$. With these definitions, the trace condition becomes simply $|\bm{t}|^2=1$, or equivalently (since we only remove zero elements to produce $\bm{\theta}$),
\begin{equation}
\label{theta_cond}
|\bm{\theta}|^2=1.
\end{equation}
That is, the Cholesky vector is of unit length, so that physical states are restricted to lying on an $(N^2-1)$-sphere in what we might call `Cholesky space'. To derive the CCRB --- the Cramer-Rao bound that incorporates this equality constraint --- we need to evaluate the Fisher information using the Cholesky parameterization, which requires differentiation of the measurement statistics with respect to $\bm{\theta}$. Substituting Eq.~(\ref{cholesky}) in to Eq.~(\ref{prob1}), we find
\begin{equation}
\label{prob_chol1}
p_\ag = \textrm{tr}\left\{T\Pi_\ag T^\dagger\right\}.
\end{equation}
In terms of $\bm{t}$, we have
\begin{equation}
\label{prob_chol2}
p_\ag = \bm{t}^\mathsf{T}P_\ag \bm{t},
\end{equation}
where the $2N^2\times 2N^2$ matrices $P_\ag$ are generated from the POVM elements $\Pi_\ag$ according to the relation
\begin{equation}
\label{Ps_def}
P_\ag = M \times \left[ \begin{array}{cc} I\otimes \Pi_\ag & \\ & I\otimes \Pi_\ag^\mathsf{T} \end{array}\right] \times M^{-1},
\end{equation}
with $I$ the $N\times N$ identity matrix and $M$ the matrix that separates real and imaginary parts, defined by \cite{Van-Den-Bos:1994bl,Brandwood:1983oe}
\begin{equation}
\label{M_def}
M = \left[\begin{array}{cc} 1 & 1 \\ -\mi & \mi \end{array}\right]\otimes I_{N^2},
\end{equation}
where $I_{N^2}=I\otimes I$ is the $N^2\times N^2$ identity matrix. Finally, we arrive at the following bilinear expression for the statistics in terms of $\bm{\theta}$
\begin{equation}
\label{prob_chol3}
p_\ag = \bm{\theta}^\mathsf{T}Q_\ag \bm{\theta},
\end{equation}
where the $N^2\times N^2$ matrices $Q_\ag$ are formed from the $P_\ag$ by deleting rows and columns with indices given by those of the elements of $\bm{t}$ that are deleted to produce $\bm{\theta}$. The $Q_\ag$ mirror the properties of the POVM elements themselves, being real, symmetric matrices that sum to the identity operator,
\begin{equation}
\label{Q_props}
Q_\ag = Q_\ag^\mathsf{T}\,,\qquad \sum_\alpha Q_\ag = I_{N^2}.
\end{equation}
Differentiating the log-likelihood function with respect to $\bm{\theta}$, we find the Fisher information in the Cholesky representation to be
\begin{eqnarray}
\nonumber
F &=&  4\sum_\ag \frac{\lambda_\gamma}{p_\ag}Q_\ag \bm{\theta}\bm{\theta}^\mathsf{T}Q_\ag\\
\label{fisher_cholesky} &=& Z^\mathsf{T}\Lambda P^{-1} Z,
\end{eqnarray}
where the diagonal matrices $\Lambda$ and $P$ are as defined in Section \ref{section:algorithm}, and where the rows of the $n_\mathrm{tot}\times N^2$ matrix $Z$ are given by the row vectors
\begin{equation}
\label{Z_build}
\bm{z}_\ag = 2\bm{\theta}^\mathsf{T}Q_\ag.
\end{equation}
Note that unlike the matrix $A$ appearing in the Bloch representation (see Eq.~(\ref{fisher_prod})), $Z$ depends on the state $\bm{\theta}$, because the state enters the measurement statistics quadratically, rather than linearly, in the Cholesky representation. This more complicated dependence on the state makes further manipulations, such as averaging over the state space, more involved than in the Bloch representation. But we can find the CCRB, and we can find the associated OED numerically.

\subsection{The constrained Cramer-Rao bound}
\label{subsection:CCRB}
In general, a set of $n$ equality constraints on $\bm{\theta}$ can be written as $\bm{s}(\bm{\theta})=0$, where $\bm{s}$ is an $n$-dimensional column vector of constraint functions. Let the derivative of $\bm{s}$ with respect to $\bm{\theta}^\mathsf{T}$ be the $n\times N^2$ matrix $S$. The CCRB is then given by \cite{Stoica:1998uk,Jagannatham:2004fe,Gorman:1990eq}
\begin{equation}
\label{CCRB}
B_C = \textrm{tr}\left\{U\left(U^\mathsf{T}FU\right)^{-1}U^\mathsf{T}\right\},
\end{equation}
where $U$ is the unitary matrix whose columns span the null space of $S$,
\begin{equation}
\label{null_U}
SU=0.
\end{equation}
Since $U^\mathsf{T}U = I_{N^2}$, the cyclic property of the trace gives
\begin{equation}
\label{CCRB_2}
B_C = \textrm{tr}\left\{\left(U^\mathsf{T}FU\right)^{-1}\right\}.
\end{equation}
In our case, this formula produces an extremely simple result. We have just a single equality constraint, so that $\bm{s}=s=|\bm{\theta}|^2-1$. The matrix $S$ then collapses to the row vector $2\bm{\theta}^\mathsf{T}$, and then $U$ can be any matrix whose columns span the $(N^2-1)$-dimensional subspace that is orthogonal to $\bm{\theta}$. Now, multiplying out the product $F\bm{\theta}$ reveals that $F\bm{\theta} = 4\bm{\theta}$, so that $\bm{\theta}$ is always an eigenvector of the Fisher matrix with eigenvalue $4$. Since $F$ is Hermitian, its $N^2-1$ remaining eigenvectors $\bm{u}_j$ are all orthogonal to $\bm{\theta}$, and to each other, and we can write
\begin{equation}
\label{F_eig}
F = 4\bm{\theta}\bm{\theta}^\mathsf{T} + \sum_{j=2}^{N^2} F_j \bm{u}_j\bm{u}_j^\mathsf{T},
\end{equation}
where the $F_j$ are the eigenvalues of $F$ associated with the $\bm{u}_j$. Now, we are free to take the $\bm{u}_j$ as the columns of $U$, and then we have
\begin{eqnarray}
\nonumber U^\mathsf{T}FU &=& 4U^\mathsf{T}\bm{\theta}\bm{\theta}^\mathsf{T}U + \sum_{j=2}^{N^2} F_j U\bm{u}_j\bm{u}_j^\mathsf{T}U,\\
\nonumber &=& \left(\begin{array}{cccc} F_2 & \phantom{0} & \phantom{0} & \phantom{0}\\
\phantom{0} & F_3 & \phantom{0} & \phantom{0}\\
\phantom{0} & \phantom{0} & \ddots & \phantom{0} \\
\phantom{0} & \phantom{0} & \phantom{0} & F_{N^2}\end{array}\right).
\end{eqnarray}
Inverting this diagonal matrix is trivial, and so we obtain the final expression for the CCRB
\begin{equation}
\label{CCRB_reduces}
B_C = \sum_{j=2}^{N^2} \frac{1}{F_j} = \textrm{tr}\left\{F^{-1}\right\}-\frac{1}{4}.
\end{equation}
That is, up to an additive constant, the CCRB is the same as the unconstrained CRB in the Cholesky representation.

\begin{figure}[h]
\begin{center}
\includegraphics[width=\columnwidth]{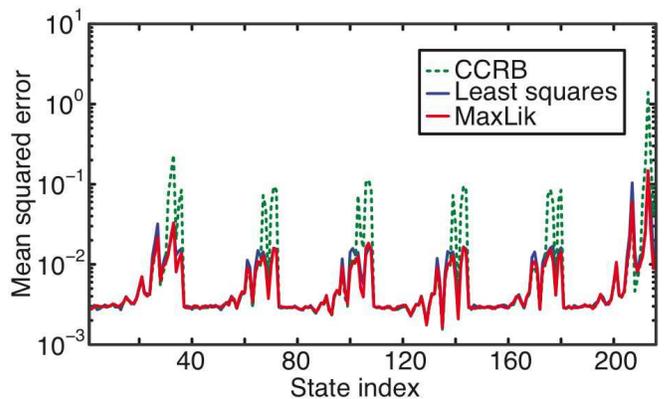}
\caption{The mean squared error $\langle | \bm{\theta}-\hat{\bm{\theta}}|^2\rangle$ achieved by the maximum likelihood and least squares methods is plotted alongside the CCRB for the same set of sample states as used to generate Figure \ref{figure:monte_carlo_compare}. In these Monte-Carlo simulations, we chose a random minimal set of binary qubit POVMs, and a uniform experiment design. We then averaged the errors over $2000$ virtual experiments, involving $10^5$ measurements each --- it seems that convergence on the CCRB requires larger samples than for the CRB in the Bloch representation. The precision attained by both constrained estimation methods is clearly well-described by the CCRB, although it is discernible that maximum likelihood performs slightly better, as might be expected. For some states, the CCRB seems to `blow up', predicting much worse performance than actually achieved. These states are close to the pure state boundary, on which the Fisher matrix in the Cholesky representation becomes singular.}
\label{figure:monte_carlo_cholesky}
\end{center}
\end{figure}

In Figure \ref{figure:monte_carlo_cholesky}, we show the results of another Monte-Carlo simulation, in which we compare the mean squared errors $\langle | \bm{\theta}-\hat{\bm{\theta}}|^2\rangle$ predicted by the CCRB with those produced by maximum likelihood estimation, and by the least squares method, for the same set of qubit states as depicted in part (a) of Figure \ref{figure:states}. The CCRB describes the errors achieved by both methods well, although the least squares method is slightly less precise. It is not possible to compare the errors produced by inversion, since non-positive density matrices cannot be represented with a Cholesky vector. 

For pure states, the Cholesky matrix $T$ contains just a single non-zero diagonal element, so that the Cholesky vector $\bm{\theta}$ contains zero elements. The determinant of the outer product $\bm{\theta}\bm{\theta}^\mathsf{T}$ then vanishes, rendering the Fisher information singular. This explains the divergences of the CCRB for pure states; of course the precision actually achieved remains finite! Numerical failure in computing the CCRB is easily avoided by slightly shifting the pure states away from the boundary of the Bloch sphere.

\subsection{Experiment designs for constrained estimators}
\label{subsection:constrained}
The CCRB can be used to find the OED for constrained estimators. The analytic formulae from Section \ref{section:analytic} do not quite carry over to the Cholesky representation, but the numerical method described in Section \ref{section:algorithm} can be applied directly. Eq.~(\ref{fisher_cholesky}) for the Fisher information in the Cholesky representation has precisely the same structure as Eq.~(\ref{fisher_prod}) for $F$ in the Bloch representation. Ignoring the constant shift of $\tfrac{1}{4}$ in Eq.~(\ref{CCRB_reduces}), finding the OED simply requires that we minimize the trace of the inverse of $F$. To find the solution, we implement exactly the same numerical algorithm as for the Bloch representation, the only difference being that we replace the matrix $A$ with $Z$ in the Eqs.~(\ref{DiffJ})---(\ref{Kdef}).

To examine the difference between these two optimizations --- the first appropriate to unbiased tomography and the second designed for constrained estimators --- we would like to compare the average OEDs predicted by the two calculations. Unfortunately we cannot use any of the techniques described previously to evaluate the average OED in the Cholesky representation, since the Fisher information defies attempts to average or invert it analytically. But it is possible to arrive at the average OED --- or at least \emph{an} average OED --- by brute force. First, one calculates the OED for each of the sample states $\bm{r}_j$ shown in part (a) of Figure \ref{figure:states}, so as to obtain $\bm{\lambda}^\mathrm{OED}(\bm{r}_j)$. Next, one uses polynomial interpolation to numerically approximate the average of $\bm{\lambda}^\mathrm{OED}$ over the Bloch sphere, yielding the design
\begin{equation}
\label{Brute}
\bm{\lambda}^{\langle\mathrm{OED:brute}\rangle} =\frac{1}{{\frac{4}{3}\pi R_2^3}} \int \bm{\lambda}^\mathrm{OED}(\bm{r})\,\rmd V,
\end{equation}
which is literally the average OED. This method is only feasible for qubit states, where a direct numerical average can be performed quickly by polynomial interpolation with a few sample points. When this brute force method is implemented in the Bloch representation, we have found that the resulting distributions seem to coincide very closely with the designs $\bm{\lambda}^{\langle \mathrm{OED}\rangle}$ and $\bm{\lambda}^{\dubav{ \mathrm{OED}}}$ introduced in Section \ref{section:average}; this is why we still refer to $\bm{\lambda}^{\langle\mathrm{OED:brute}\rangle}$ as the average OED.

We compared the average OEDs for the Bloch and Cholesky representations using $1000$ randomly chosen minimal sets of binary qubit POVMs. For each POVM set, we calculated the `discrepancy' $\mathcal{D}$ as the sum of the absolute difference between the two designs,
\begin{equation}
\label{discrepancy}
\mathcal{D}=\sum_\gamma \left|\lambda^{\langle\mathrm{OED:brute}\rangle}_{\gamma,\mathrm{Bloch}}-\lambda^{\langle\mathrm{OED:brute}\rangle}_{\gamma,\mathrm{Cholesky}}\right|.
\end{equation}
Since by definition the $\lambda$'s sum to unity, one would expect an average discrepancy of order $1$, when comparing two completely unrelated distributions. We found $\langle \mathcal{D}\rangle\sim 0.0418(7)$ for the average discrepancy, which shows that the optimal design for reconstruction by inversion is very close to the optimal design for maximum likelihood tomography. Although we are not able to compare the ODTs for the two parameterizations (because the Cholesky representation is too unwieldy), this suggests that the results we derived in the Bloch representation for unbiased estimators are useful for constrained estimators too, even though the Bloch representation does not account for the constraints.

Despite this, the precision of maximum likelihood estimation seems to improve less than the precision of inversion upon adoption of the average OED (whether derived using either the Bloch or the Cholesky representation), indicating that this reconstruction technique is less amenable to optimization generally. It seems that maximum likelihood estimation is able to recruit the positivity constraint to counteract the statistical errors that a poor experiment design exacerbates. After our foray into the Cholesky representation, we are able to conclude that the smaller improvements in maximum likelihood tomography are `intrinsic' to the method, and are not due to the neglect of the positivity constraint when optimizing the design in the Bloch representation. Of course, if precision matters, it is always better to use an optimized design over a uniform one.

\section{Conclusion}
\label{section:conclusion}
We have revisited the problem of optimal experiment design for quantum state tomography, first introduced in \cite{Kosut:2004pi}. We have shown that specialist convex optimization software is not necessary to calculate the OED, and further that analytic results exist for minimal tomography, in which the measurements do not overdetermine the state. The reliance of the OED on knowledge of the true state is rather paradoxical, but we have explored a number of averaging methods that remove this state dependence. The averaging relies on approximating the state space as hyperspherical (an exact procedure for qubits), but the results are generally insensitive to the radius chosen for the hypersphere, indicating that this approximation is reasonably robust.

We introduced the average OED, which optimizes the average precision, or the `precision on average', depending on the details of the method --- the resulting designs appear to be the same. We have also considered the ODT, which seeks to render a tomographic experiment maximally fair. Fortunately, it appears that the average OED and the ODT are rather similar, so that fair measurements are generally precise. Finally, we confirmed that the formalism correctly describes the achieved precision for unbiased state estimation techniques, but that constrained estimators appear to perform better than predicted. However a new formulation of the problem allows the treatment of these constrained estimators, and shows that designs optimal for unbiased estimators are also very close to optimal for constrained methods.

Although this paper contains a rather large number of results, it seems that many of the various optimal designs we have proposed are effectively interchangeable. One can use the ODT, or any one of three methods for computing the average OED (if one includes the brute force method for qubits), or one can use the Cholesky representation. Numerical evidence suggests there is little to choose between them. This is good news for experimentalists who would like an answer to the question ``What experiment design is optimal in quantum tomography, when I don't already know what state I expect?'' After all is said and done, probably the most useful response is ``The numerical method for computing the average OED described in Section \ref{subsection:averaging_fish}.'' This method is fast, and works for any number of measurements on a system of arbitrary dimension \footnote{Details of the simple (though not pretty) Matlab codes used are available from JN on request.}.

Further work would help to justify some of the claims we have made, particularly regarding the applicability of our averaging method in higher dimensions. An interesting possibility is to apply these techniques in examining the effectiveness of a detector; perhaps they can be used to design better measurements under laboratory constraints. In any case, as quantum tomography becomes increasingly indispensable, we hope that our results will prove useful in real experiments.

\acknowledgements
JN wishes to thank R. Kosut for his valuable advice on some preliminary results.
This work was supported by the EPSRC through the QIP IRC
(GR/S82716/01) and project EP/C51933/01. JN
thanks Hewlett-Packard. IAW was
supported in part by the European Commission under the Integrated
Project Qubit Applications (QAP) funded by the IST directorate as
Contract Number 015848, and the Royal Society.

\bibliography{../references/references}
\end{document}